# Long-term stability and temperature variability of Iris AO segmented MEMS deformable mirrors

M.A. Helmbrecht [*a], M. He [a], C.J. Kempf [a], F. Marchis [a,b]
[a]Iris AO, Inc., 2930 Shattuck Ave. #304, Berkeley, CA 94705-1775, USA;
[b]SETI Institute, 189 Bernardo Ave., Suite #100, Mountain View, CA 94043, USA.

## ABSTRACT

Long-term stability of deformable mirrors (DM) is a critical performance requirement for instruments requiring open-loop corrections. The effects of temperature changes in the DM performance are equally critical for such instruments. This paper investigates the long-term stability of three different Iris AO PTT111 DMs that were calibrated at different times ranging from 13 months to nearly 29 months prior to subsequent testing. Performance testing showed that only a small increase in positioning errors occurred from the initial calibration date to the test dates. The increases in errors ranged from as little as 1.38 nm *rms* after 18 months to 5.68 nm *rms* after 29 months. The paper also studies the effects of small temperature changes, up to 6.2°C around room temperature. For three different arrays, the errors ranged from 0.62-1.42 nm *rms*/°C. Removing the effects of packaging shows that errors are ≤0.50 nm *rms*/°C. Finally, measured data showed that individual segments deformed ≤0.11 nm *rms*/°C when heated.

## 1. INTRODUCTION

A number of deformable mirror (DM) technologies are suitable for astronomical instrumentation and adaptive optics systems. One class of DMs, electrostatically actuated microelectromechanical systems (MEMS) based DMs, offer high-order corrections with no hysteresis, stable operation, and fast temporal response. The highly repeatable electrostatic actuators used with MEMS DMs dissipate negligible power, thus self heating is not an issue as it can be with magnetically actuated and piezoelectric actuated DMs. Further, the materials typically used to fabricate MEMS DMs are very stable. As this paper shows, stability for the Iris AO PTT111 segmented MEMS DM is excellent over timescales of years. For comparison, Table 1 summarizes test results from the literature for a few other DMs. From this table, it is clear that MEMS DMs are far superior for applications requiring stable operation.

Table 1: Comparison of DM position stability from the literature and the work here.

| Reference | DM Type | Actuation Method | Creep/Drift (nm *rms*) | Timeframe (days) | Creep/Drift (nm *rms*/hr) |
|---|---|---|---|---|---|
| Vdovin *et al.* [1] | Membrane | Electrostatic | 31.7 | 16 | 0.083 |
| Bitenc *et al.* [2] | Membrane | Magnetic | 80<br>15 (drift after creep has settled) | 0.25<br>4.7 | 13.3<br>0.133 |
| Morzinski *et al.* [3] | MEMS Facesheet | Electrostatic | No quantitative data reported: "flattening voltages continue to give a sharp internal PSF" | 1,500 | NA |
| Iris AO PTT111 | MEMS Segment | Electrostatic | 1.38 | 558 | 0.0001 |
| Iris AO PTT111 | MEMS Segment | Electrostatic | 5.68 | 868 | 0.00027 |

[*] michael.helmbrecht@irisao.com

a) b) c)

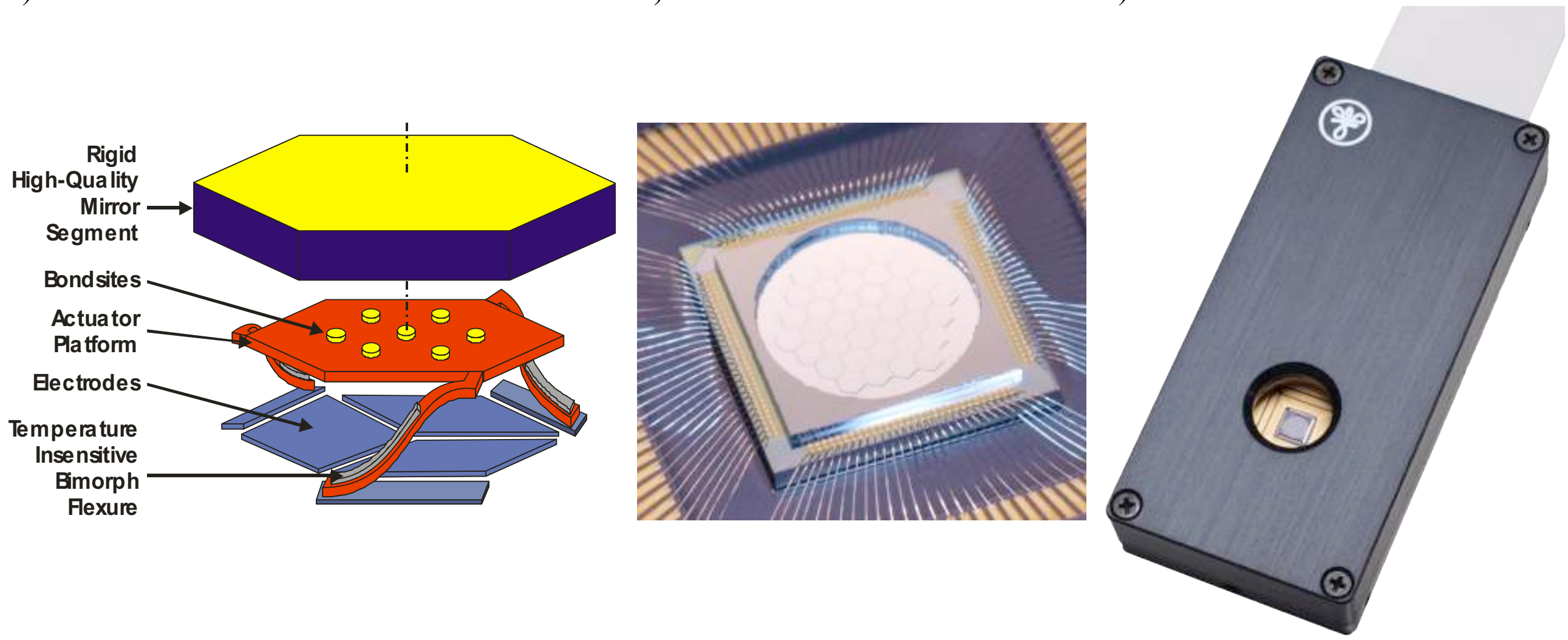

Figure 1: a) Exploded-view schematic diagram of the 700 µm diameter (vertex-to-vertex) mirror segment. The mirror segment is capable of moving in piston/tip/tilt motions (PTT). Scaling is highly exaggerated in the vertical direction. 37 of these segments are tiled to create the DM array in b). Die photograph of a 111-actuator 37-piston/tip/tilt-segment DM with 3.5 mm inscribed aperture (Product name: PTT111). This DM is packaged into the mechanical housing in c). The MEMS DM is mounted onto a ceramic package and sealed in a nitrogen environment behind a cover window. The ceramic package is soldered to a printed circuit board mounted inside of the mechanical enclosure shown here.

## 2. IRIS AO DM BACKGROUND

Figure 1a is a schematic diagram of one of the 37 piston/tip/tilt (PTT) segments in the PTT111 DM shown in Figure 1b. The DM is manufactured using typical MEMS and integrated circuit materials such as polycrystalline silicon (polysilicon), silicon dioxides, silicon nitrides, and a proprietary bimorph material with similar coefficient of thermal expansion (CTE) to that of polysilicon [4]. The s-shape of the bimorph flexures that elevates the DM segment is a result of engineered residual tensile stresses in the bimorph and actuator-platform polysilicon.

To actuate the DM, the red actuator-platform layer is held at ground potential and the three diamond-shaped electrodes are energized at different electrical potentials. Applying the same voltage to all three electrodes pulls the segment in a piston motion toward the electrodes. A differential voltage across the electrodes results in tip and tilt motions. Because the positioning is highly repeatable, the DM segment motion can be calibrated, thus linearizing the DM position into orthogonal coordinates [5].

After the DM is fabricated using highly stable MEMS materials, it is mounted onto a ceramic pin-grid array (PGA) package using an epoxy. The DM is sealed in nitrogen by epoxying a cover window over the DM. The PGA package is soldered into a printed circuit board (PCB). The PCB is then mounted into a mechanical enclosure. A fully packaged PTT111 DM in its enclosure is shown in Figure 1c. This study analyzes the stability and temperature effects of DMs packaged as shown Figure 1c. It thus mimics the performance of DMs used in the field.

## 3. TEST AND ANALYSIS PROCEDURES

For this study, three different DMs were randomly chosen that had been calibrated at different dates. Information for the DMs is shown in Table 2 including coating type and segment thickness. Two of the DMs were fully packaged as shown in Figure 1c. The third DM (FSC37-01-09-1211) was mounted in an older package where the dust cover glass is mounted to the enclosure lid instead of to the DM package itself.

All DM surface measurements presented here were made with the Zygo NewView 7300 white-light interferometer that Iris AO uses for DM calibrations. The microscope-style interferometer requires that the DMs be mounted with the optical surface horizontal. The DMs were affixed to a motorized X-Y stage using a standard mounting bracket provided with the DM. The drive electronics used to operate the mirror were positioned within 0.5 m of the DM and a temperature sensor.

Table 2: DM information for the stability and temperature testing described herein. *This DM used older packaging that did not include a sealed window.

| DM Serial Number | Coating Type | Coating Thickness (nm) | Segment Thickness (µm) | Calibration Date | Calibration Temperature (°C) |
|---|---|---|---|---|---|
| FSC37-02-01-0901 | protected-Silver (Ag) | 200 | 25 | May 11, 2015 | 19.8 |
| FSC37-01-11-0614 | bare Gold (Au) | 125 | 25 | December 10, 2014 | 21.1 |
| FSC37-01-09-1211* | protected-Aluminum (Al) | 200 | 50 | January 31, 2014 | 22.5 |

During testing, the ambient temperature was not controlled, but stayed relatively constant (<0.4°C) during a measurement set. For temperature variability testing, an air conditioner used to cool the cleanroom containing the DM, drive electronics and interferometer was turned off to increase the ambient temperature. Temperatures were recorded for all measurements. The DMs were positioned using the drive electronics they were calibrated with and the calibration file from the initial DM calibration. For all measurements shown here, the DMs were open-loop flattened by setting all of the segment PTT positions to [Piston = 0 µm, Tip = 0 mrad, Tilt = 0 mrad].

### 3.1 Data Analysis

To analyze test data, all Zygo measurements were imported into Matlab. To minimize reconstruction noise from segment edges, a 15 µm exclusion area was masked off of the segments. Prior to *rms* calculations, extreme outliers in the data ($\geq 5\sigma$) were considered as bad data points and were removed. Data showing variability from temperature are the difference of measurements taken at two temperatures. Data showing stability measurements are absolute measurements.

For all measurements, the first order Zernike modes (piston, tip, tilt) have been removed from the entire DM array. No corrections have been made to the individual mirror segment positions. The Zernike modes that have been removed are denoted in the figures and shown pictorially. Temperature and stability data are also shown with higher-order terms removed from the measured mirror array data. This is done to highlight low-order effects believed to be predominantly from packaging. Doing so also shows the benefits an offline calibration would provide to correct for drift and large temperature excursions. Finally, all data presented are for the mirror figure. The associated wavefront errors would be twofold larger when the DM used to reflect an incoming wavefront.

## 4. TEMPERATURE VARIABILITY

### 4.1 Mirror array figure errors versus temperature

The first series of tests presented in Figure 2 demonstrate the effects of a temperature change to a flattened DM surface. The tests were conducted around room temperature and spanned a range of up to 6.2°C as listed in Figure 2. The mirror-surface difference-measurement plots shown have been normalized to 1°C. With first-order Zernike terms removed, the variations in surface figure range from 0.62-1.42 nm *rms*/°C as seen in the first-row images. The DMs that have been sealed with nitrogen have smaller variations. Temperature changes cause the packages and DM to bow because of CTE mismatches between the silicon DM and the ceramic package. The DM package with the sealed window is more rigid compared to the package without it, and is thus more resilient to the effects of CTE mismatches. The bow from CTE is low order, thus removing 2$^{nd}$ order Zernike terms (focus and astigmatism) in the analysis in large part removes the effects of CTE mismatches between the package and the DM. With the CTE mismatches effectively removed, the DM array variations are very similar for the three DMs: 0.46-0.50 nm *rms*/°C.

Figure 2 also shows the temperature effects with higher order terms removed. At the extreme case, with 5$^{th}$ order terms removed, the variations in mirror surface figure are 0.36 nm *rms*/°C. These remaining errors are predominantly from variations in rigid-body segment positions that stem from manufacturing variations in the MEMS actuator platform geometry across the chip.

| Zernike Terms Removed | FSC37-02-01-0901<br>Sealed Window<br>T1=27.4°C, T2=22.2°C, ΔT=5.2°C | FSC37-01-11-0614<br>Sealed Window<br>T1=28.0°C, T2=21.8°C, ΔT=6.2°C | FSC37-01-09-1211<br>Dust Cover<br>T1=27.5°C, T2=22.4°C, ΔT=5.1°C |
|---|---|---|---|
| First Order<br>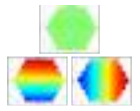 | 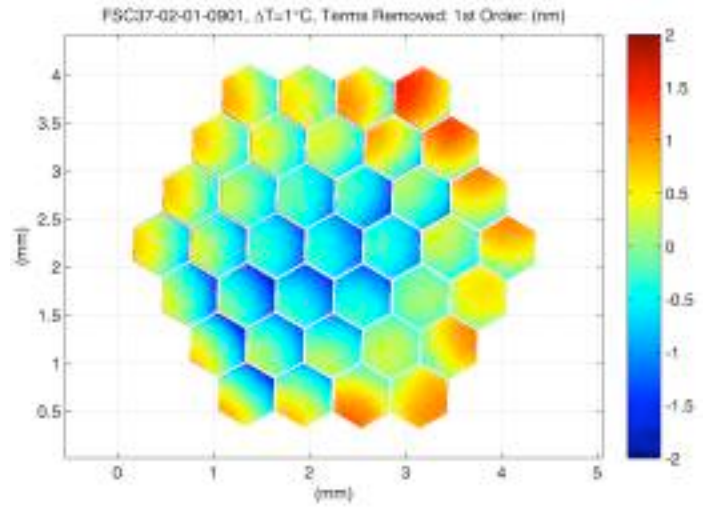<br>0.62 nm *rms*/°C | 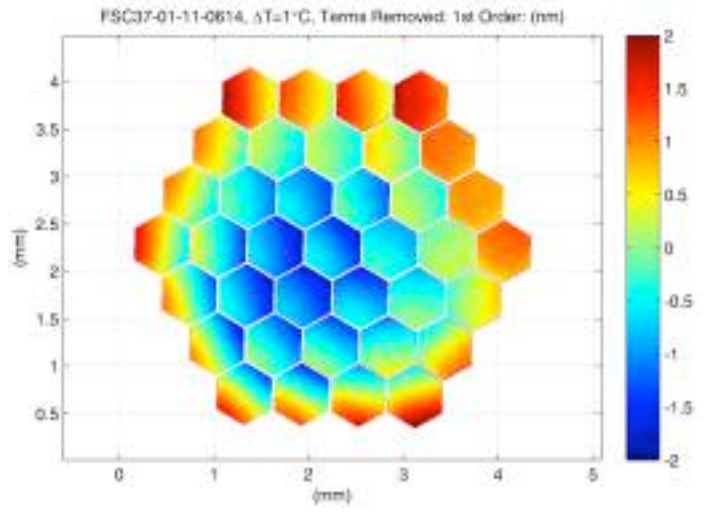<br>0.84 nm *rms*/°C | 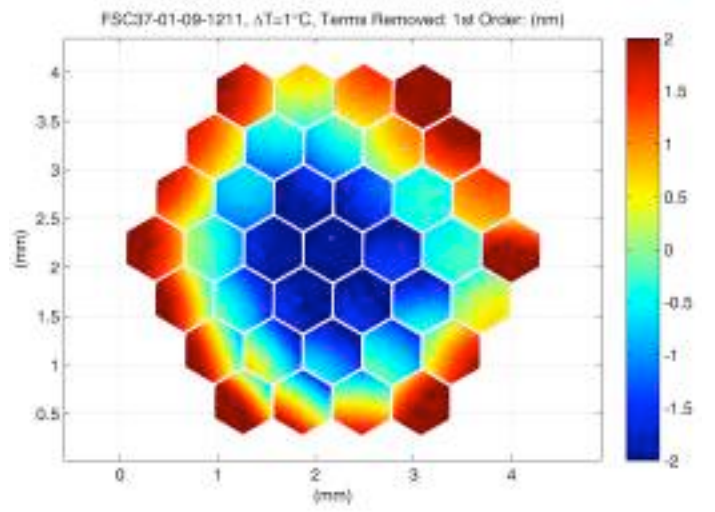<br>1.42 nm *rms*/°C |
| Second Order<br>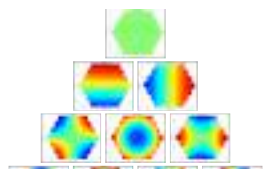 | 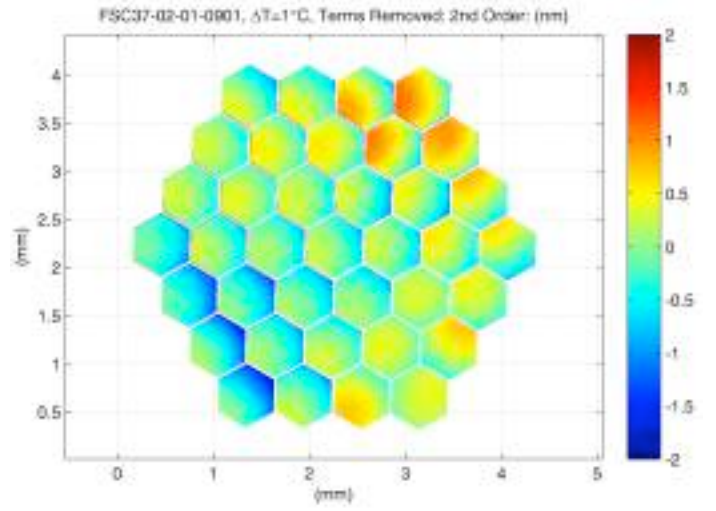<br>0.46 nm *rms*/°C | 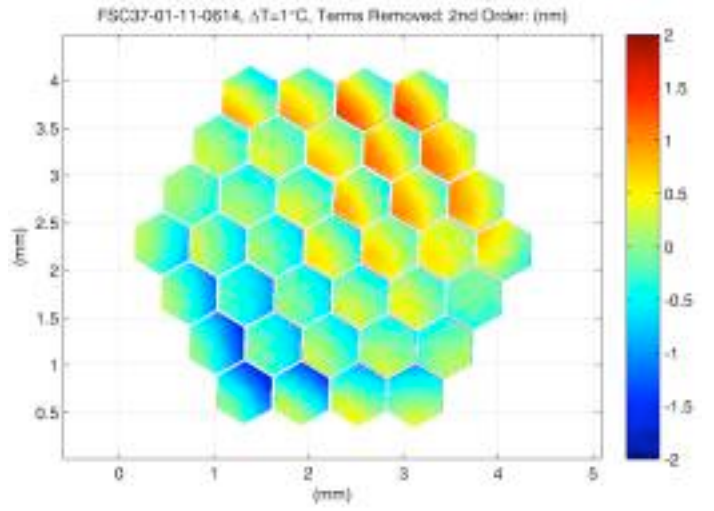<br>0.50 nm *rms*/°C | 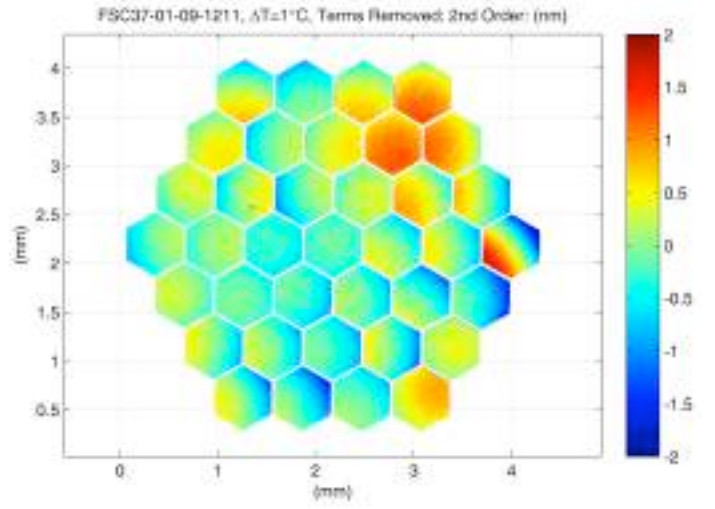<br>0.50 nm *rms*/°C |
| Third Order<br>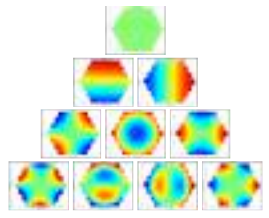 | 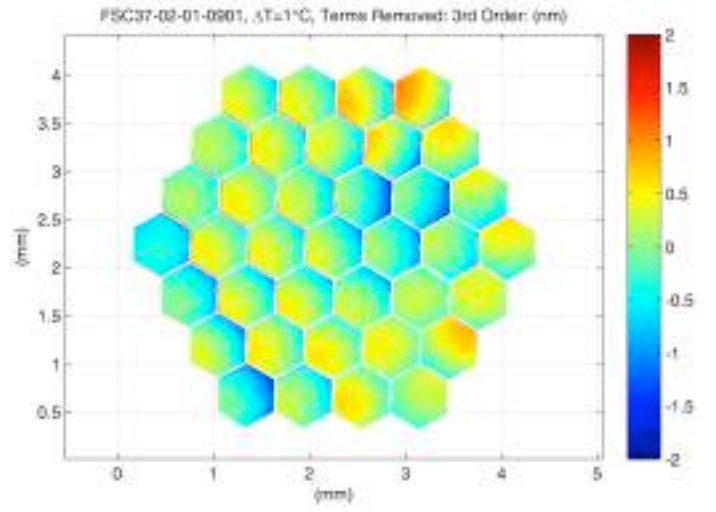<br>0.42 nm *rms*/°C | 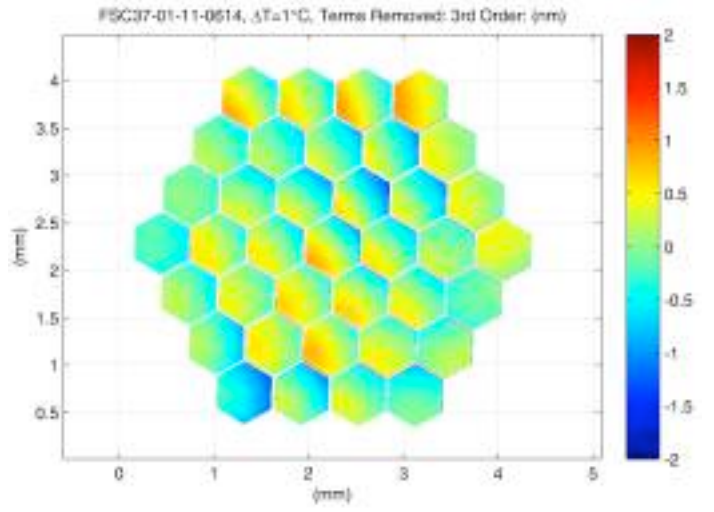<br>0.40 nm *rms*/°C | 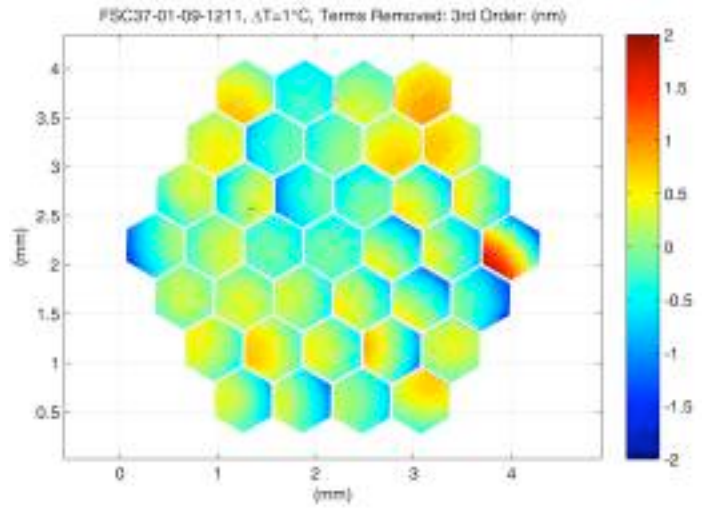<br>0.44 nm *rms*/°C |
| Fourth Order<br>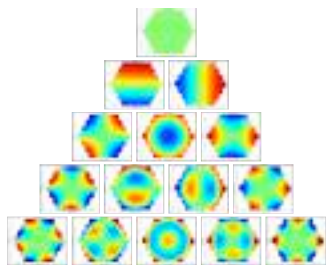 | 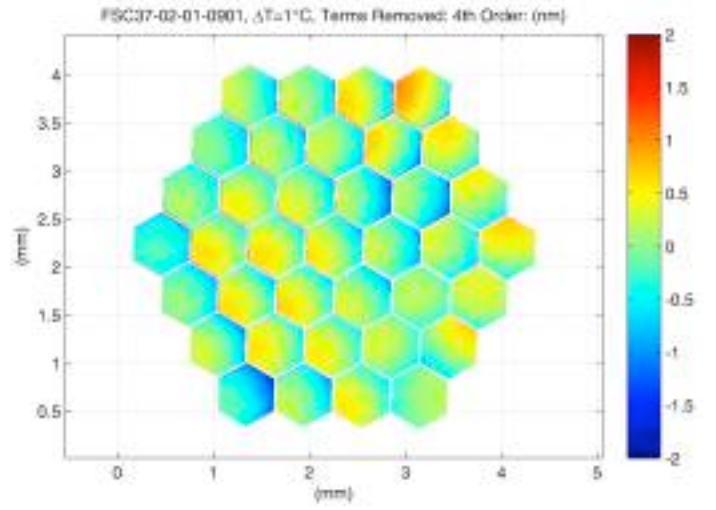<br>0.40 nm *rms*/°C | 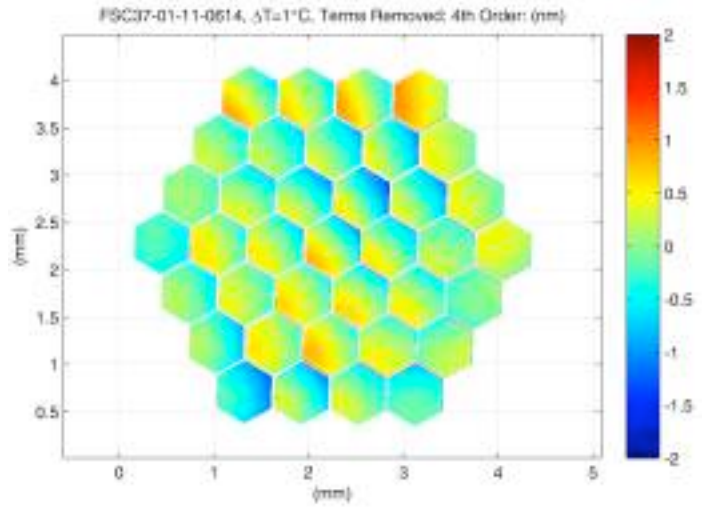<br>0.40 nm *rms*/°C | 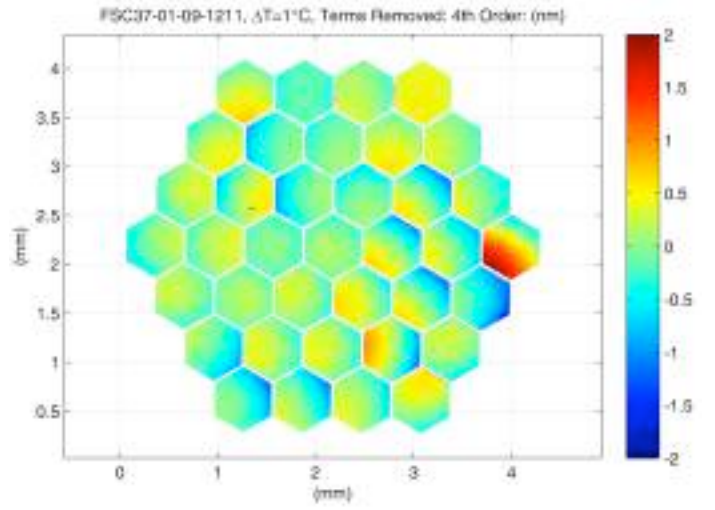<br>0.39 nm *rms*/°C |
| Fifth Order<br>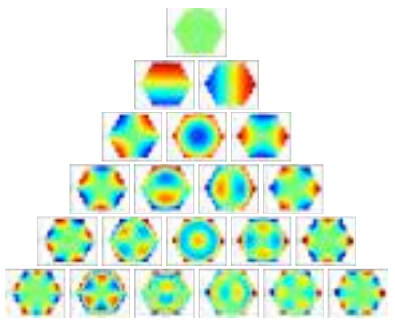 | 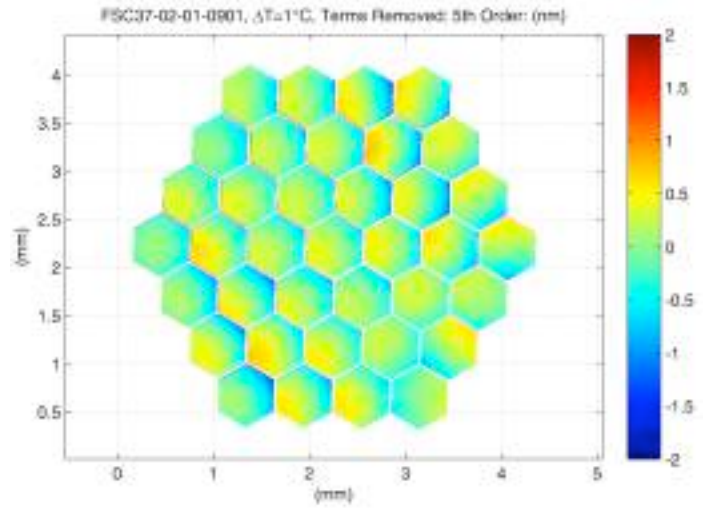<br>0.36 nm *rms*/°C | 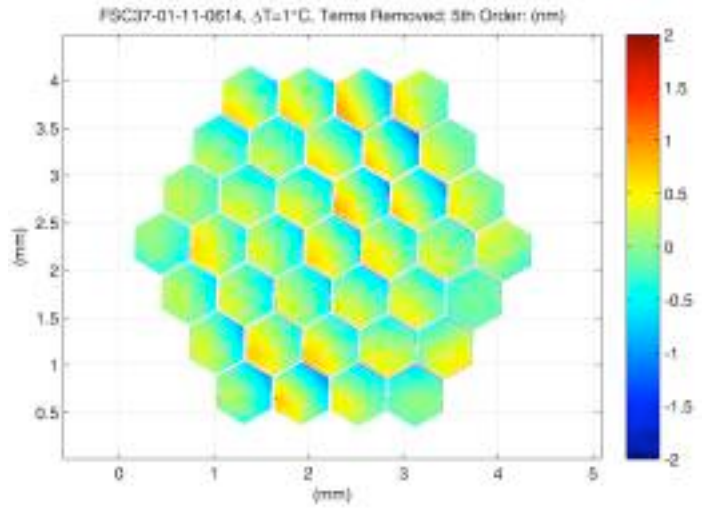<br>0.36 nm *rms*/°C | 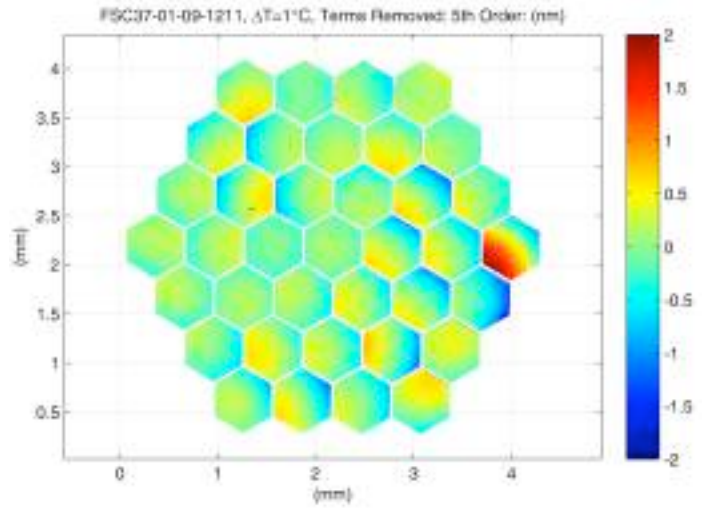<br>0.36 nm *rms*/°C |

Figure 2: Effects of temperature changes on flattened DM arrays normalized to 1°C. Color-bar limits are ±2 nm for all plots. The rows show successive orders of Zernike modes removed from the DM array measurement, starting with 1st order (piston, tip, tilt) on row to 5th order terms removal. Removing the 2nd order Zernikes terms removes most the deformation caused by CTE mismatches between the DM and package. The DMs with sealed windows have stiffer packages that deform less than the DM with the dust cover.

### 4.2 Mirror segment rigid-body motions versus temperature

Removing the piston term Figure 2 obscures the fact that the entire mirror array shifts upwards with temperature. Although the bimorph material was chosen to have a CTE that is close to polysilicon, it does not perfectly match. Thus, the average DM array height with respect to the DM substrate changes with temperature. For most applications, this effect is irrelevant. It is a pertinent effect for interferometers, however, and thus we show it here.

Figure 3 shows plots of the positions at the center of each segment for a 1°C change in temperature. These positions were extracted from the measurements used for Figure 2 and thus display the rigid body motions of the segments with temperature. No terms have been removed to compensate for the package bow. Segment numbering corresponds to the standard Iris AO numbering scheme, whereby the central segment is number 1, the first ring of segments are numbered 2-7, the second ring of segments is 8-19, and the outer ring of segments are numbered 20-37. The reference regions used for these measurements are at the periphery of the array. Given the low-order bow from CTE mismatches with the packaging, we expect the segment heights will increase from the central segment to the outer ring of segments. The 1σ noise floor for the segment height with respect to the reference regions is approximately 2.8 nm and 1σ noise floor for the segment tip/tilt positions is approximately 0.6 μrad.

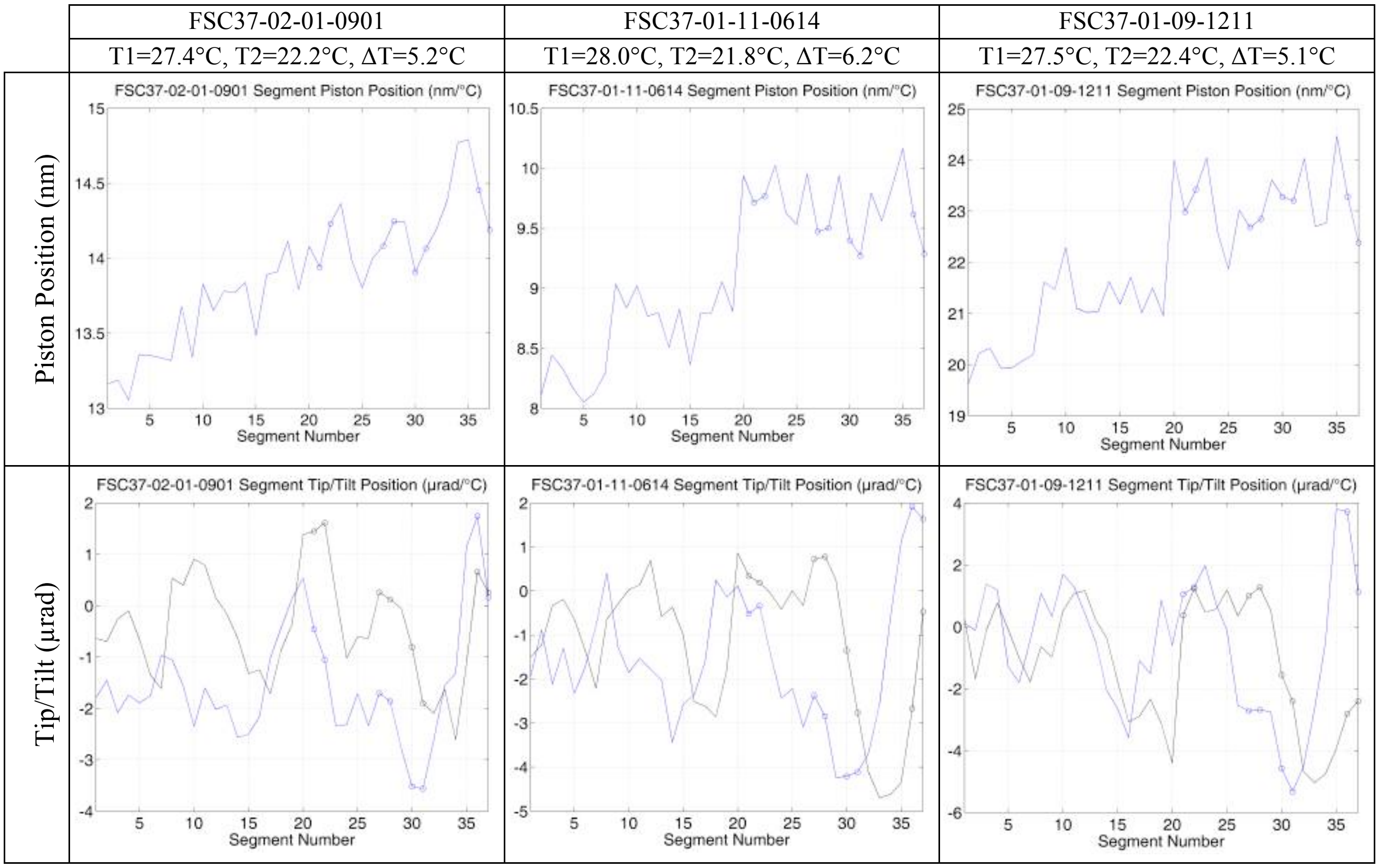

Figure 3: Segment rigid-body motions normalized to a 1°C change in temperature. The circles denote segments near the reference regions used to determine segment heights. The average piston motion shows that the entire array moves upward with temperature.

### 4.3 Mirror-segment figure versus temperature

The last temperature effect we report on is the change in surface figure versus temperature for individual segments. The DM optical coatings (protected-Silver, protected-Aluminum, and Gold) have CTEs that differ from silicon. Further, there are subtle differences in the CTEs of single-crystalline silicon and polysilicon. The net effect is the DM segments deform with temperature. The dominant effect is segment bow [6], but forces coupled in from the underlying actuator platforms that are bonded to the mirror segments will introduce additional modes. Figure 4 shows the effect of a 1°C change on the mirror segment figure. The top row of images shows the raw data and the bottom row shows the best Zernike fit for the deformation with temperature. The top rows of text show the mirror serial number, the coating type, the mirror thickness, and the temperature range for the difference measurement. For all cases, the segment bow is many times smaller than the change in DM array shape.

| FSC37-02-01-0901<br>protected-Ag, 25 µm segment | FSC37-01-11-0614<br>Au, 25 µm segment | FSC37-01-09-1211<br>protected-Al, 50 µm segment |
|---|---|---|
| T1=27.4°C, T2=22.2°C | T1=28.0°C, T2=21.8°C | T1=27.5°C, T2=22.4°C |

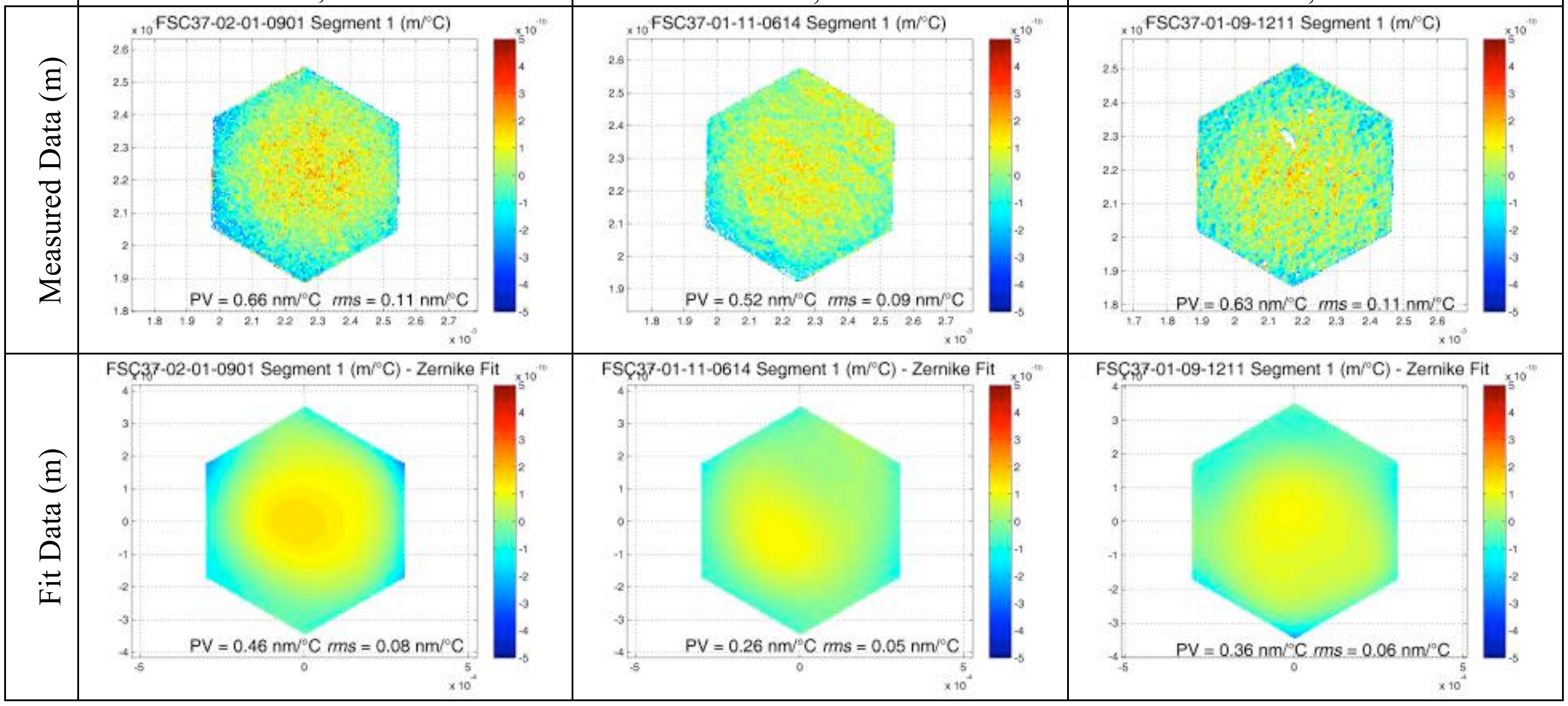

Figure 4: Effect of a 1°C temperature change on the segment figure. Color-bar limits are ±0.5 nm for all plots. (Upper row) Measured data for Segment 1 normalized to a 1°C change. (Lower row) Fit data for Segment 1 normalized to a 1°C change. Peak-to-valley (PV) and *rms* data are shown in the images.

# 5. LONG-TERM OPEN-LOOP POSITIONING STABILITY

To assess long-term calibration stability, the three DMs listed in Table 2 and their respective drive electronics were run through the same qualification tests used during the initial calibration. The temperature in the cleanroom was adjusted to be as close to the original calibration temperature as possible. At best the difference was only 0.1°C, whereas at worst the difference was 3.3°C. The data were not manipulated to compensate for temperature affects. We expect the drift shown for measurements with large temperature differences is actually less than reported because of these effects.

Figure 5-Figure 7 show the open-loop flattened DM measurements at the time of calibration and at the test date. Test dates ranged from 13 months to nearly 29 months after the initial calibration date. As with prior data, the open-loop flattened mirror shape is shown with first-order terms removed and with higher-order terms removed as well.

These tests show that even after 29 months, the calibrations hold very well. The slow drift in open-loop flattening performance ranged from as little as 1.38 nm rms to only 5.68 nm *rms* for the DM measured 29 months after calibration. Unlike with thermal effects, removing second-order terms from the measurements shows relatively little improvement for these measurements. From this we conclude that the drift in the calibrations is related to small changes in the electronics and DM actuators instead of changes in the packaging.

Most AO systems and instruments have some means to measure and calibrate residual errors. Thus the small changes in the open-loop segment positions DM could be compensated for by adding a modal correction to the DM shape using modal positioning functions provided by Iris AO. The data here show that removing $5^{th}$ order terms reduces the drift in open-loop positioning to a scant 0.71 nm *rms* after the original calibration date.

| Zernike Terms Removed | FSC37-02-01-0901<br>Calibration Date: May 11, 2015<br>T=19.8°C | FSC37-02-01-0901<br>Test Date: June 17, 2016<br>T=22.6°C |
| --- | --- | --- |
| First Order<br>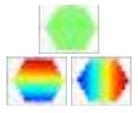 | 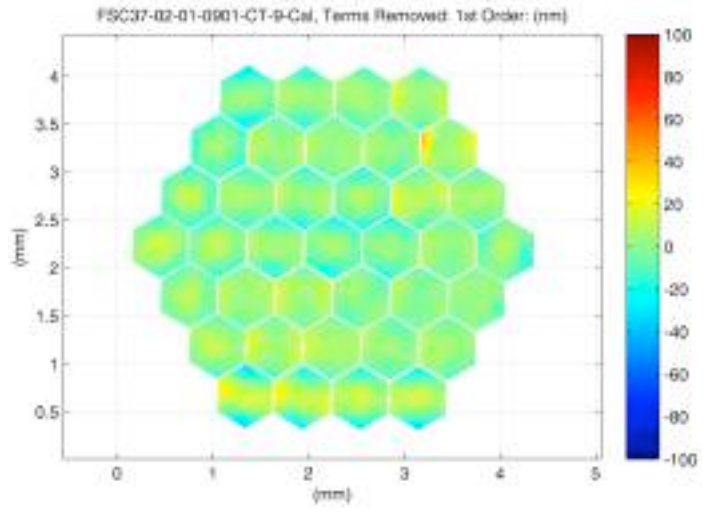<br>7.92 nm *rms* | 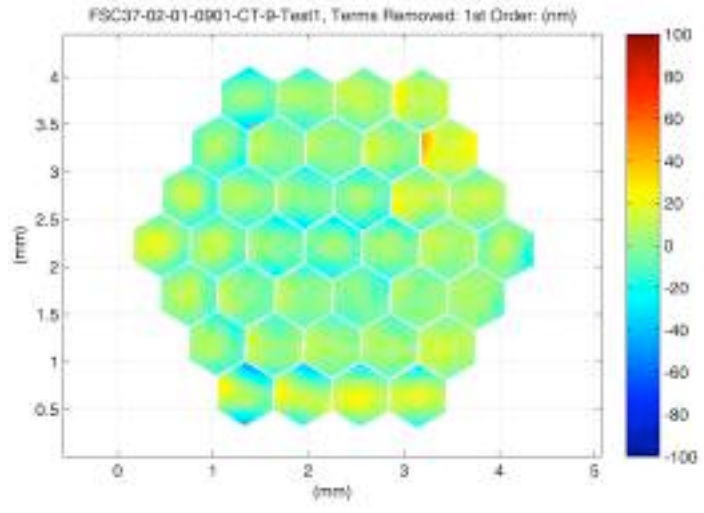<br>10.18 nm *rms* |
| Second Order<br>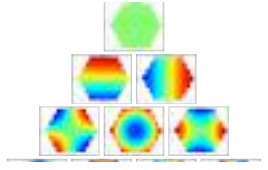 | 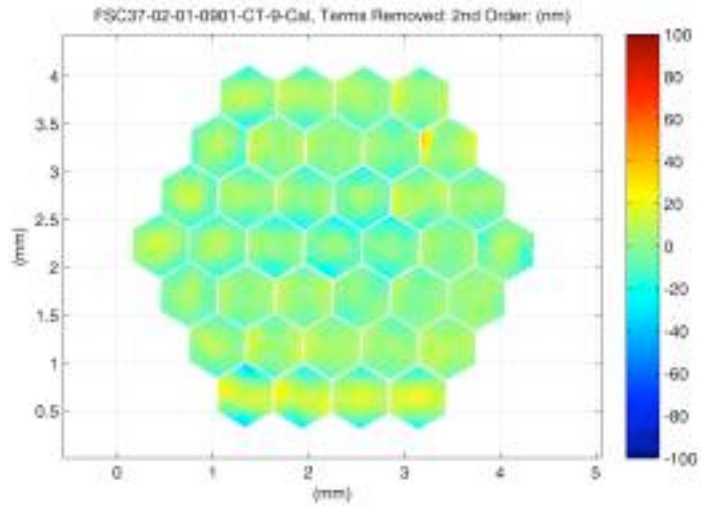<br>7.97 nm *rms* | 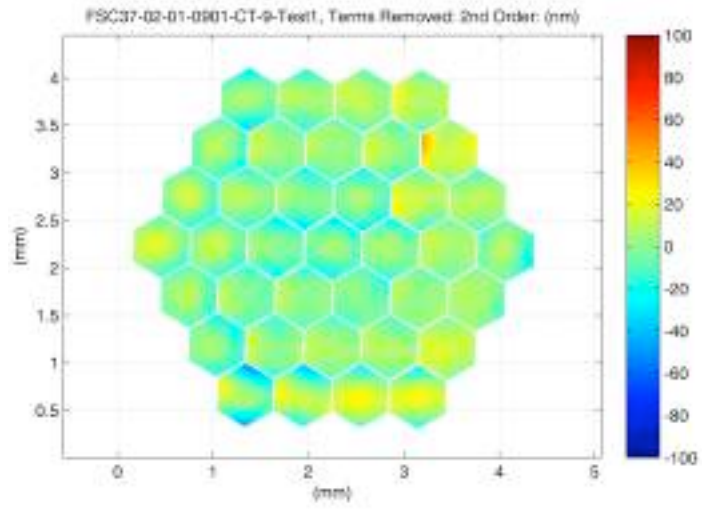<br>10.04 nm *rms* |
| Third Order<br>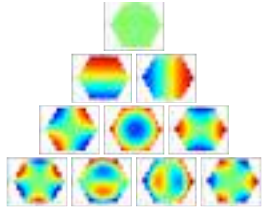 | 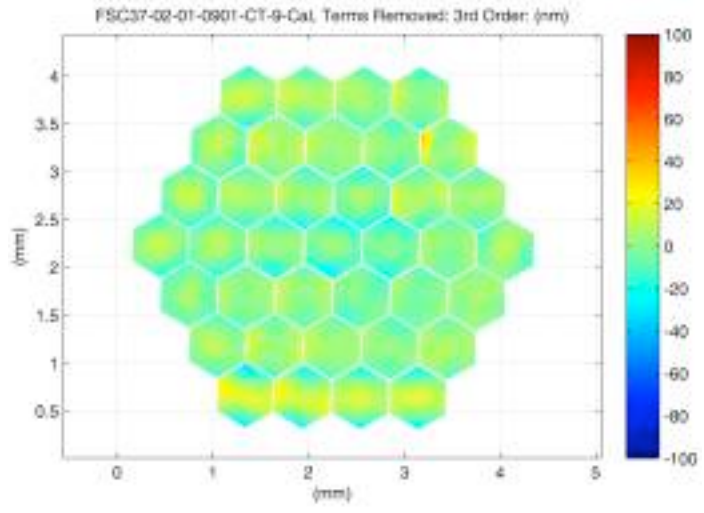<br>7.94 nm *rms* | 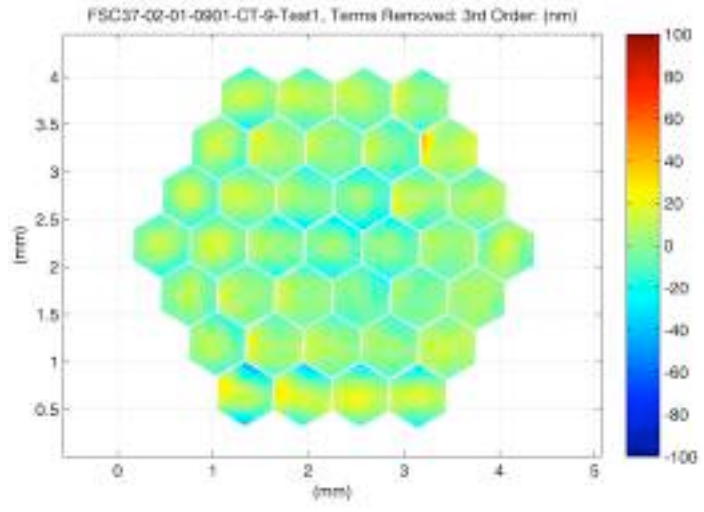<br>9.41 nm *rms* |
| Fourth Order<br>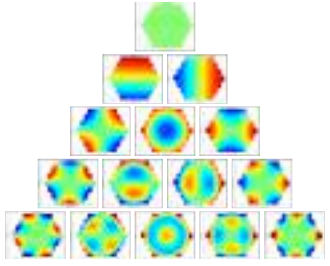 | 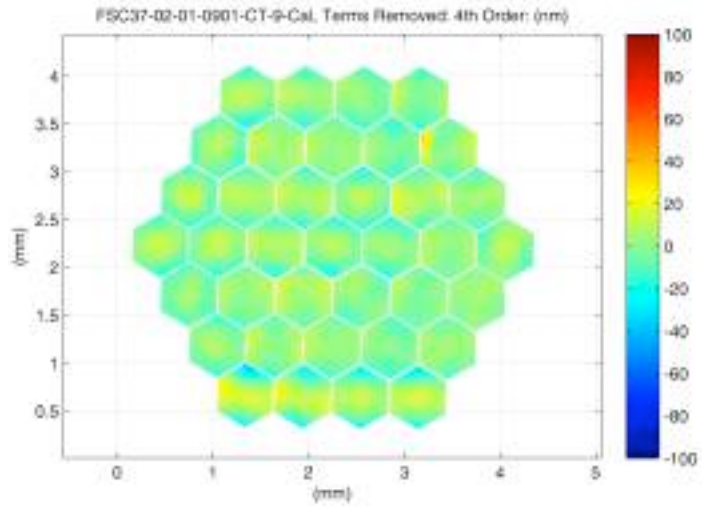<br>7.49 nm *rms* | 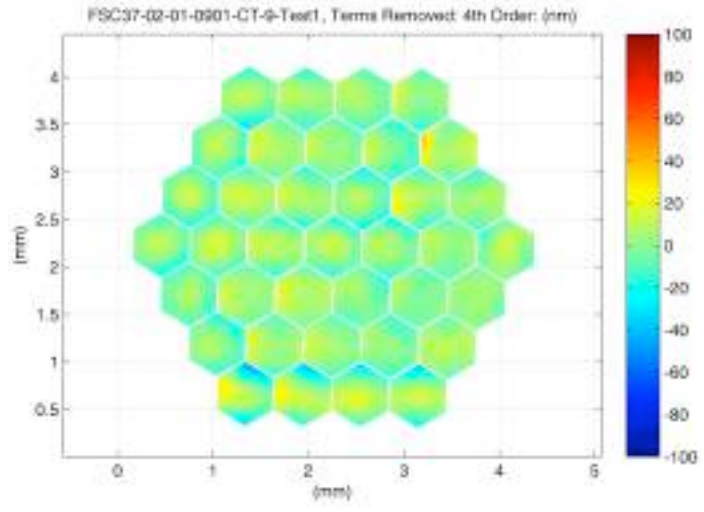<br>8.86 nm *rms* |
| Fifth Order<br>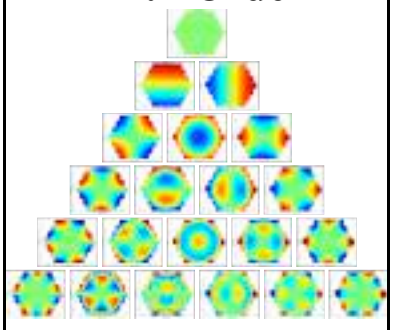 | 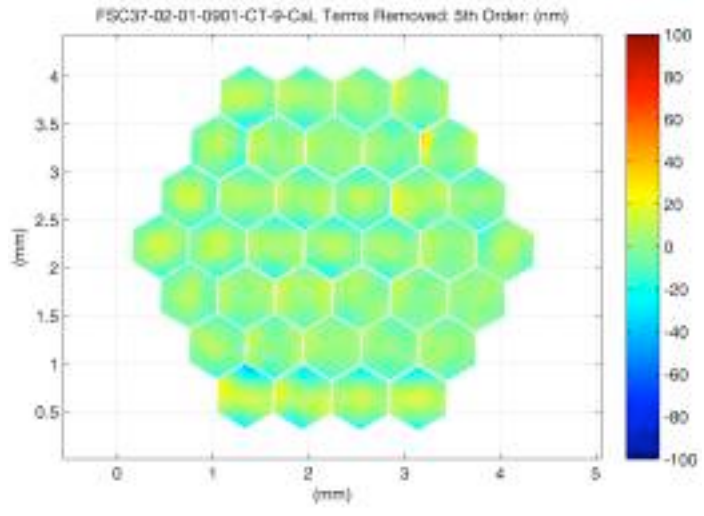<br>7.31 nm *rms* | 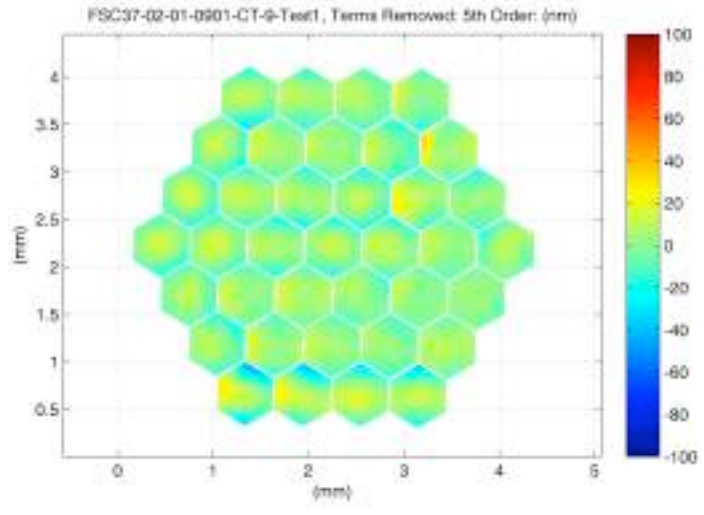<br>8.68 nm *rms* |

Figure 5: Measurements of an open-loop flattened DM at the time of calibration and 13 months after. Color-bar limits are ±100 nm for all plots. The rows show measurements with successive orders of Zernike modes removed from the DM, starting with 1st order (piston, tip, tilt) on row to 5th order terms removal.

| Zernike Terms Removed | FSC37-01-11-0614<br>Calibration Date: December 10, 2014<br>T=21.1°C | FSC37-01-11-0614<br>Test Date: June 20, 2016<br>T=21.8°C |
|---|---|---|
| First Order<br>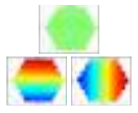 | 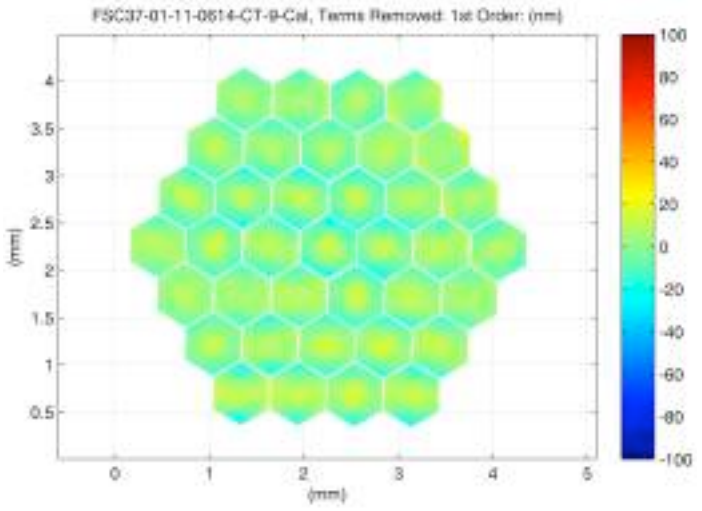<br><br>7.23 nm *rms* | 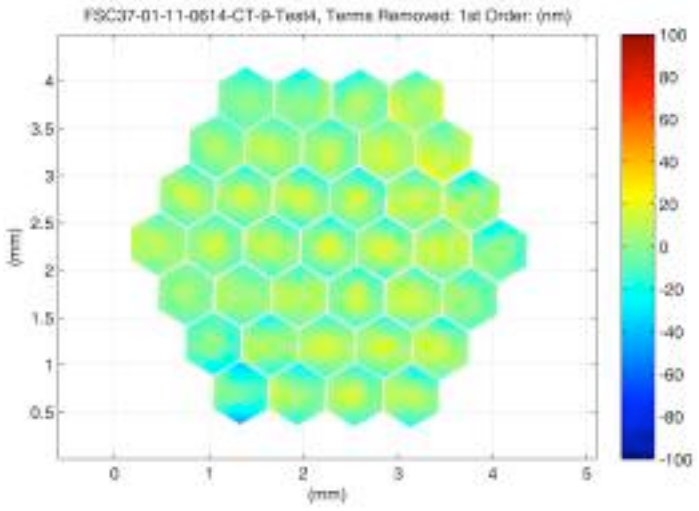<br><br>8.61nm *rms* |
| Second Order<br>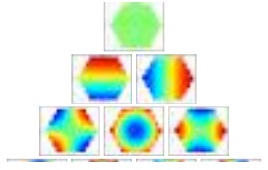 | 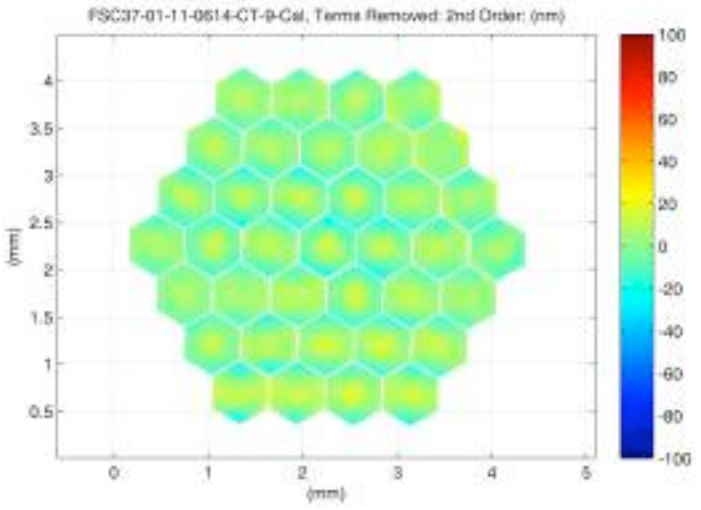<br><br>7.25 nm *rms* | 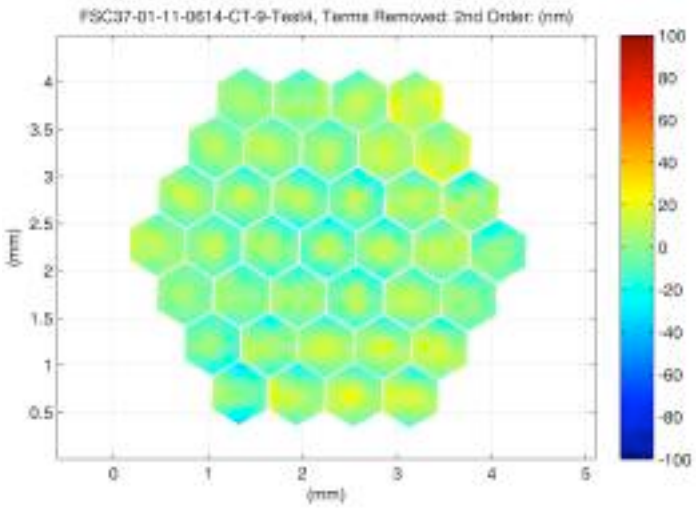<br><br>8.54 nm *rms* |
| Third Order<br>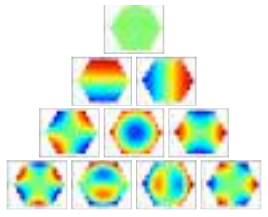 | 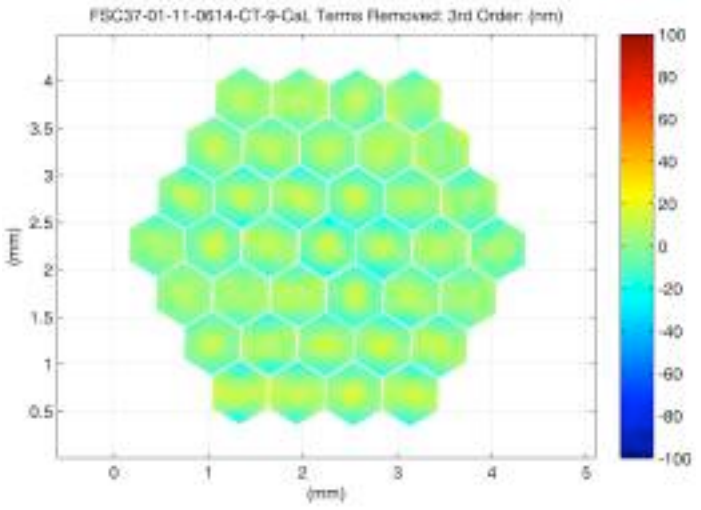<br><br>7.20 nm *rms* | 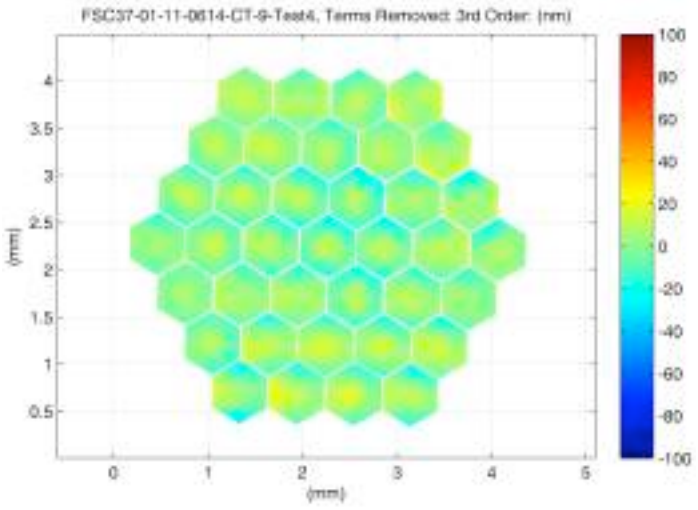<br><br>7.97 nm *rms* |
| Fourth Order<br>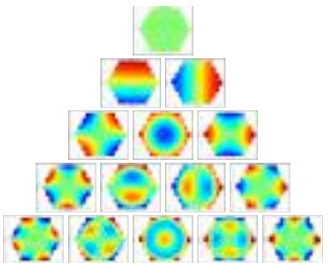 | 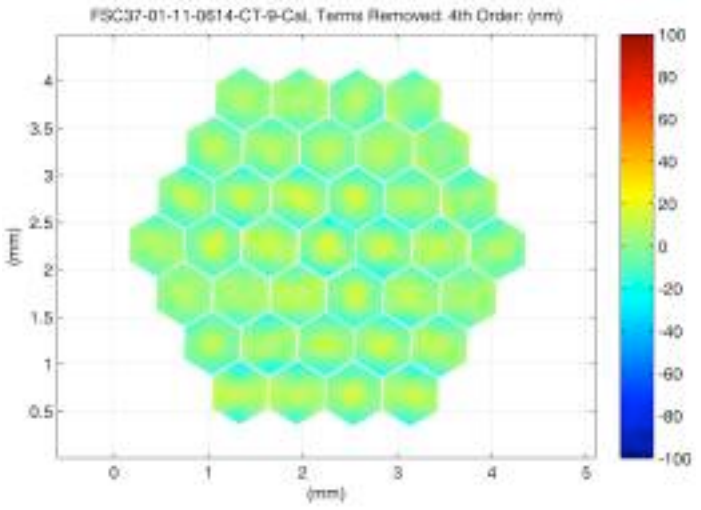<br><br>7.04 nm *rms* | 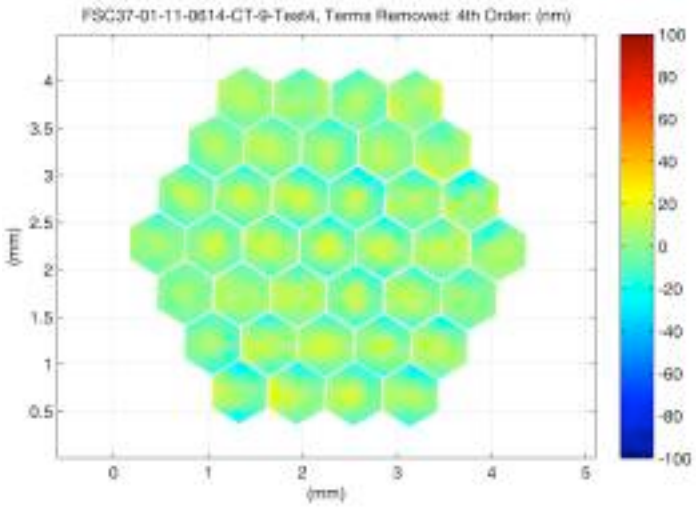<br><br>7.68 nm *rms* |
| Fifth Order<br>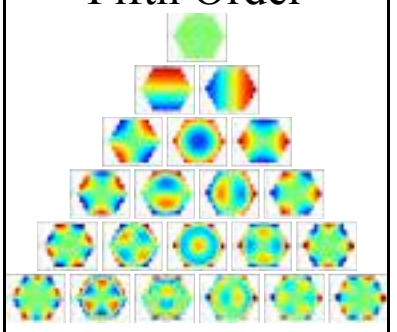 | 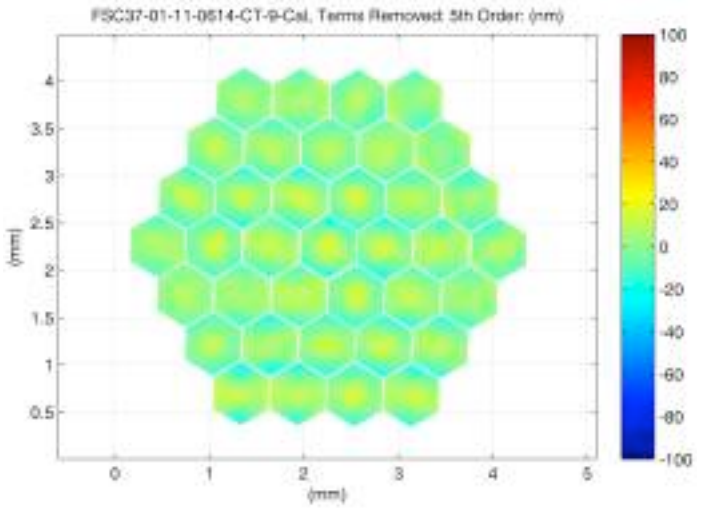<br><br>7.00 nm *rms* | 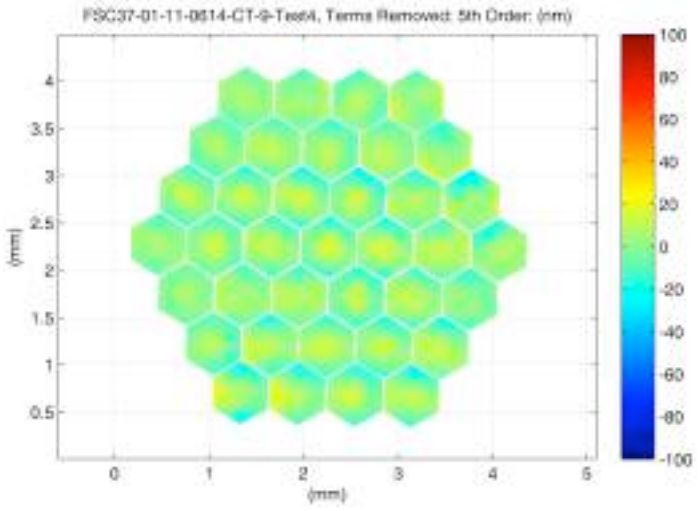<br><br>7.54 nm *rms* |

Figure 6: Measurements of an open-loop flattened DM at the time of calibration and 18 months after. Color-bar limits are ±100 nm for all plots. The rows show measurements with successive orders of Zernike modes removed from the DM, starting with 1$^{st}$ order (piston, tip, tilt) on row to 5$^{th}$ order terms removal.

| Zernike Terms Removed | FSC37-01-09-1211<br>Calibration Date: January 31, 2014<br>T=22.5°C | FSC37-01-09-1211<br>Test Date: November 20, 2015<br>T=19.2°C | FSC37-01-09-1211<br>Test Date: June 17, 2016<br>T=22.4°C |
|---|---|---|---|
| First Order<br>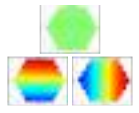 | 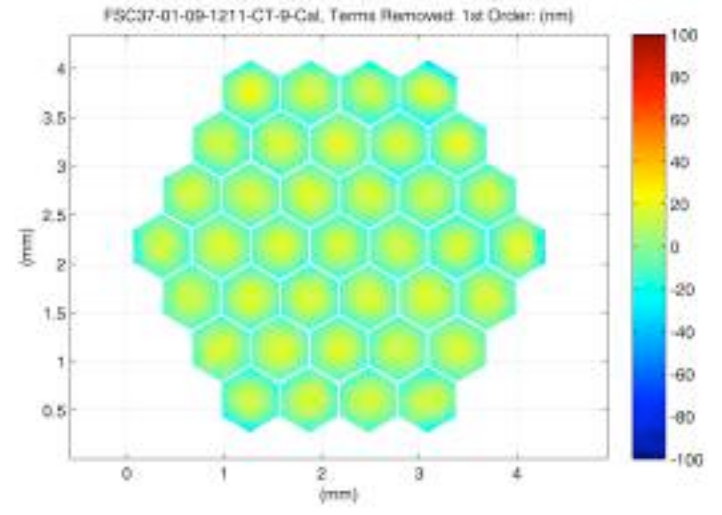<br>9.87 nm *rms* | 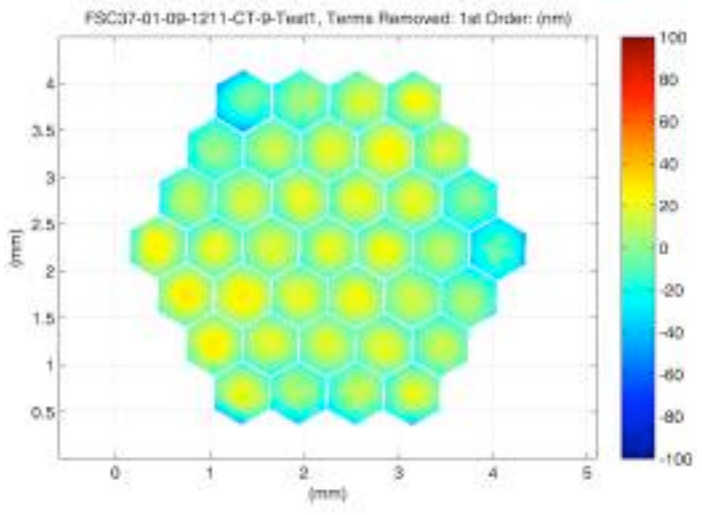<br>12.91 nm *rms* | 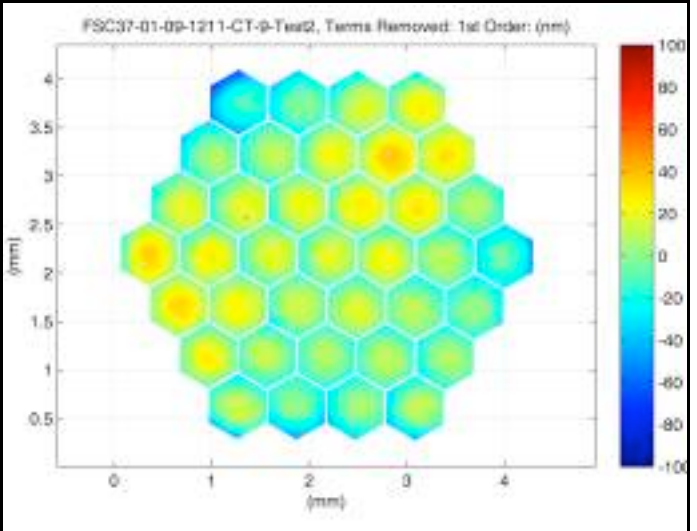<br>15.55 nm *rms* |
| Second Order<br>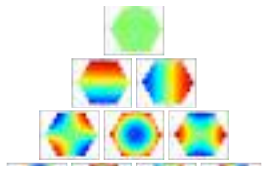 | 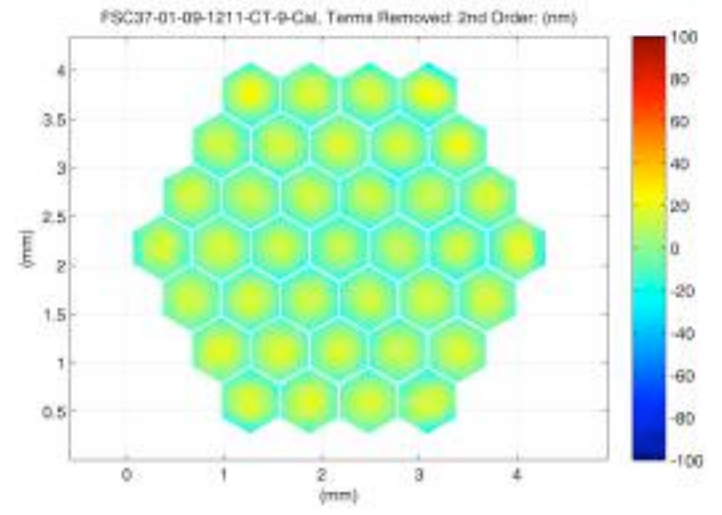<br>9.85 nm *rms* | 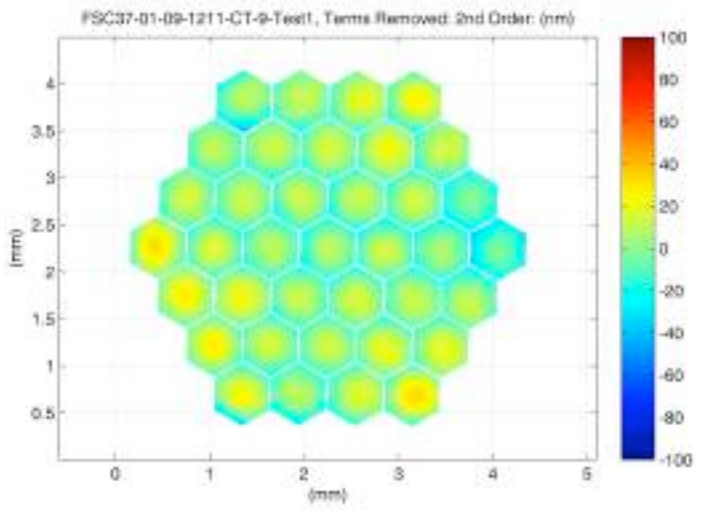<br>11.90 nm *rms* | 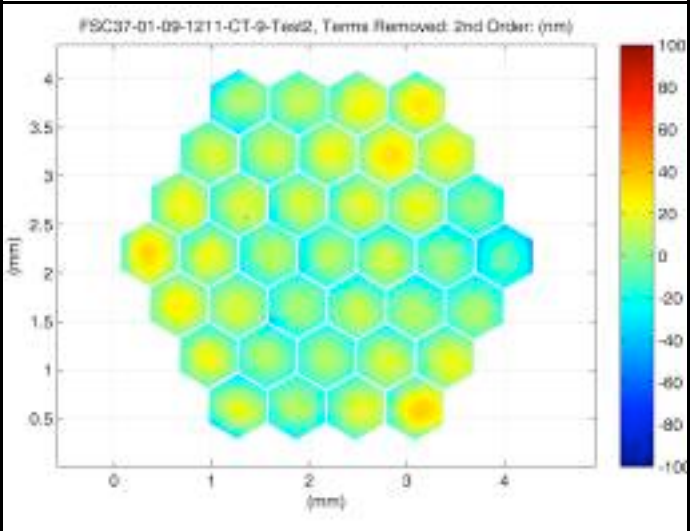<br>13.41 nm *rms* |
| Third Order<br>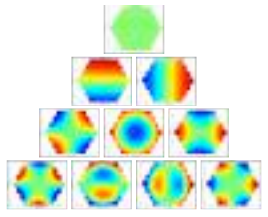 | 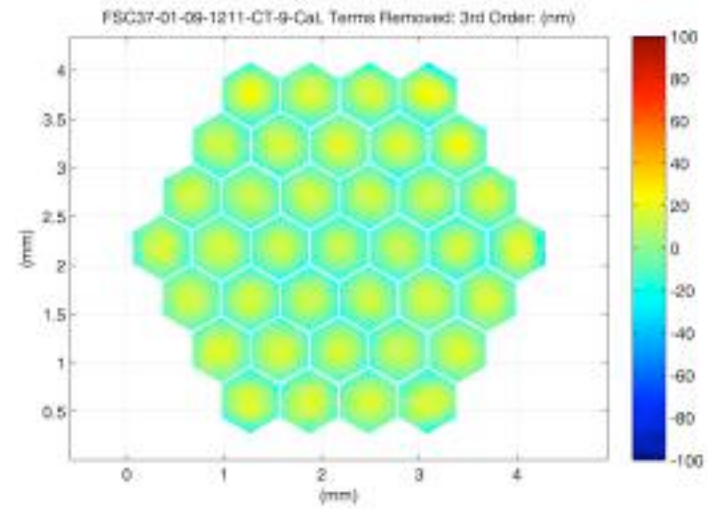<br>9.84 nm *rms* | 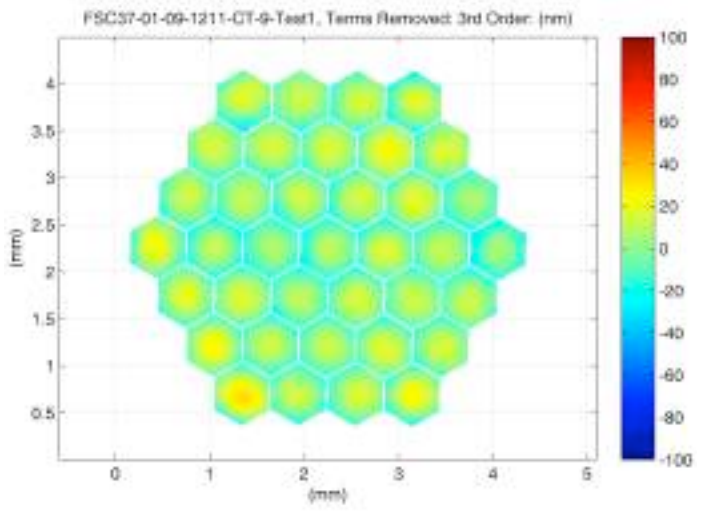<br>10.83 nm *rms* | 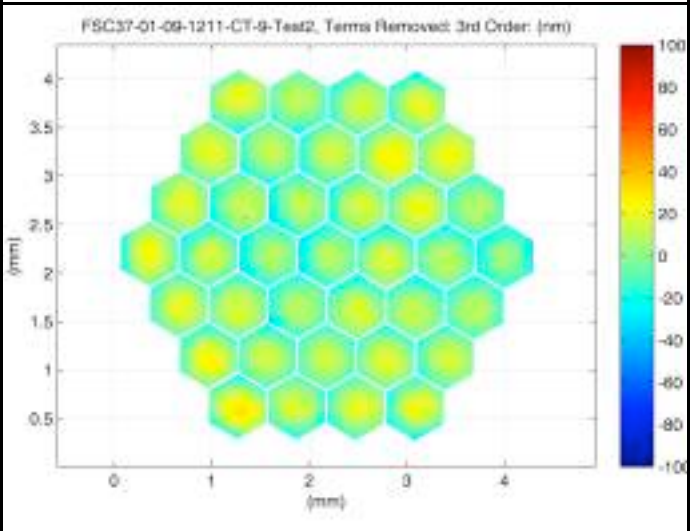<br>11.55 nm *rms* |
| Fourth Order<br>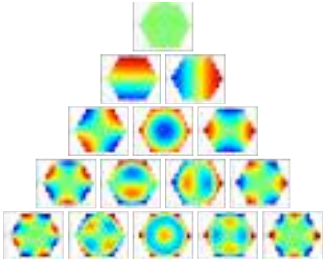 | 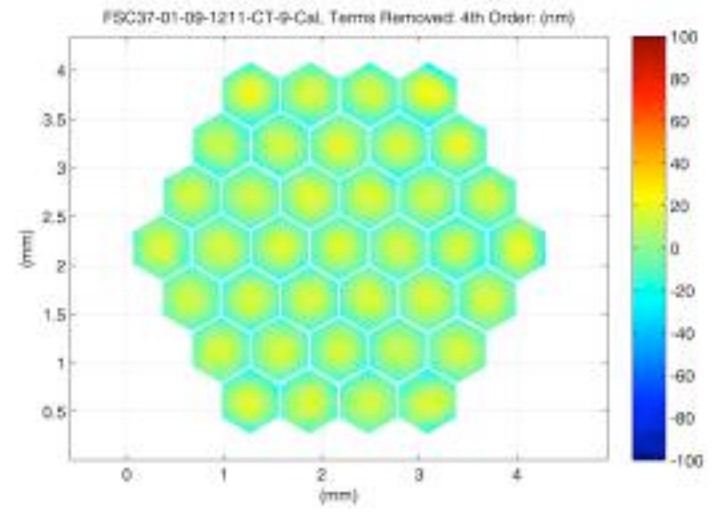<br>9.70 nm *rms* | 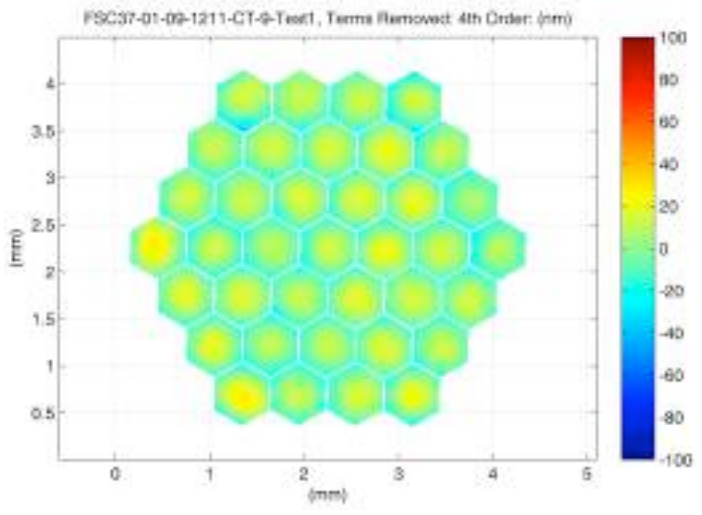<br>10.46 nm *rms* | 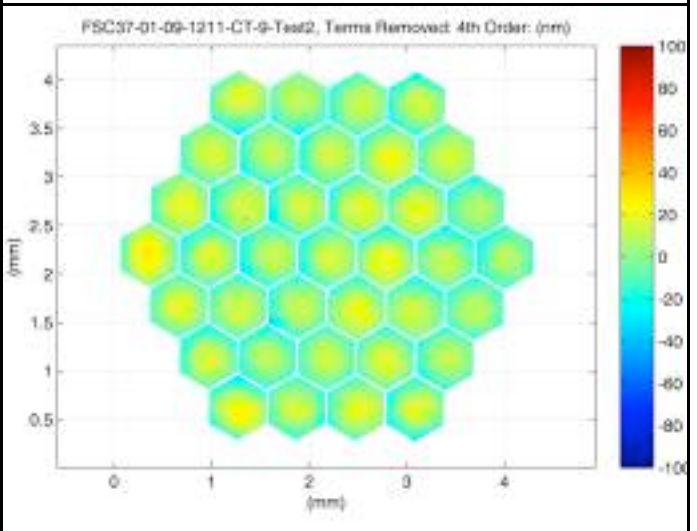<br>10.97 nm *rms* |
| Fifth Order<br>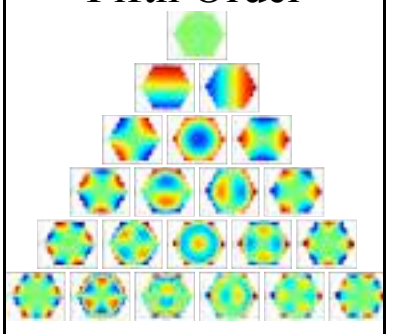 | 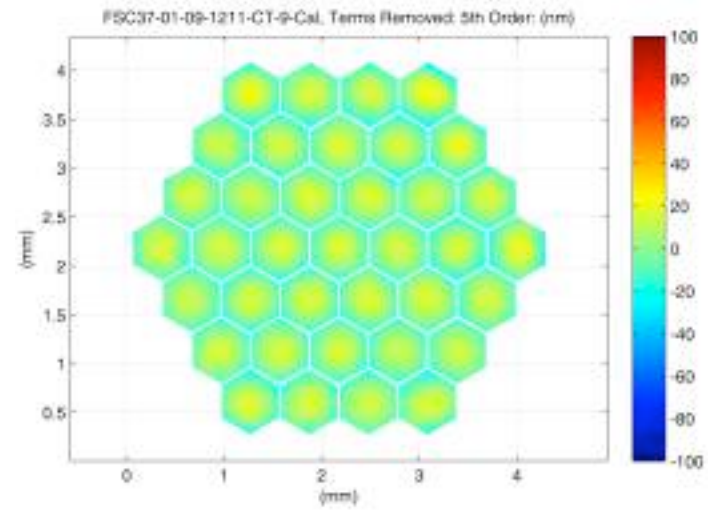<br>9.64 nm *rms* | 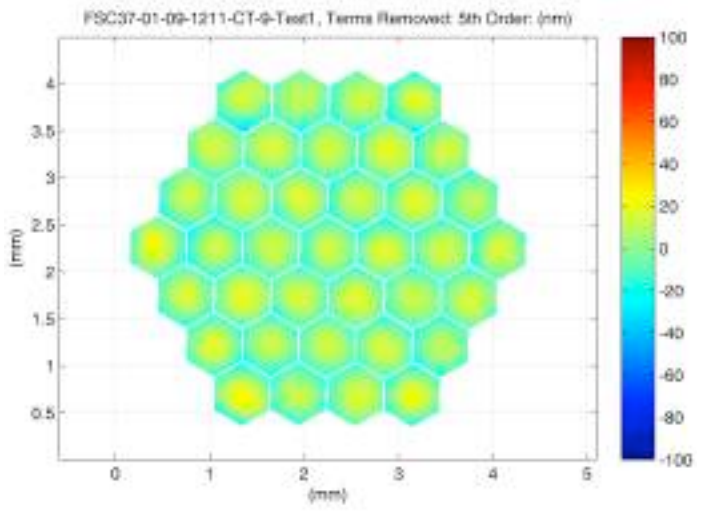<br>10.16 nm *rms* | 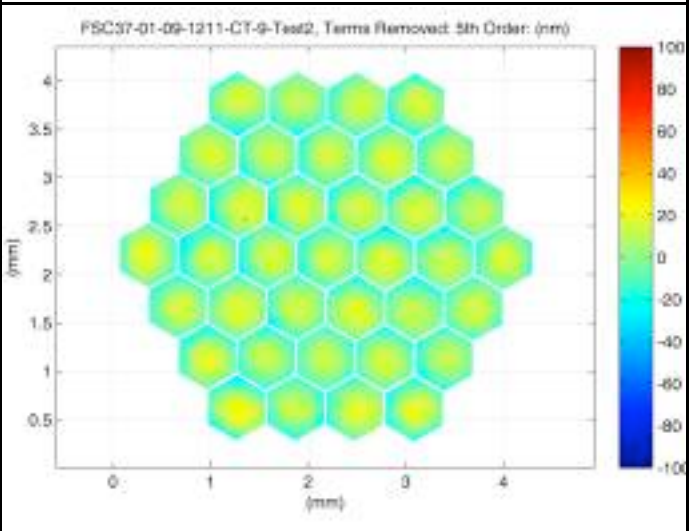<br>10.58 nm *rms* |

Figure 7: Measurements of an open-loop flattened DM at the time of calibration, nearly 22 months after, and nearly 29 months after. Color-bar limits are ±100 nm for all plots. The rows show measurements with successive orders of Zernike modes removed from the DM, starting with $1^{st}$ order (piston, tip, tilt) on row to $5^{th}$ order terms removal.

# 6. CONCLUSIONS

Measurements of open-loop-flattened DMs taken 13, 18, 22, and 29 months after the initial calibration show that the PTT111 MEMS segmented DM technology is very stable. For the worst case, open-loop flattening errors increased only 5.68 nm *rms* from 9.87 nm *rms* to 15.55 nm *rms* after 29 months. Much of this drift could be eliminated by applying a Zernike modal correction to the DM. Correcting simply for focus and astigmatism reduces the drift to 3.54 nm *rms* whereas a fifth-order correction reduces the drift to a mere 0.71 nm *rms*.

The tests conducted here also quantify the effects of temperature variations on the PTT111 DMs. Measurements show that the change in a flattened DM ranges from 0.62-1.42 nm *rms*/°C. The DM with the largest temperature variation was from an older design that is more susceptible to deformation from CTE mismatches between the MEMS DM chip and the ceramic package it is mounted on. After removing second-order Zernike terms from the analysis to reduce the effects of packaging, the DMs arrays showed temperature variability of 0.46-0.50 nm *rms*/°C. The deformation of individual mirror segments is much lower. For the three DMs measured here with Al, Ag, and Au coatings, the segments deformed only 0.05-0.08 nm *rms*/°C as determined by fitting the difference measurements with a $5^{th}$ order Zernike to reduce measurement noise. The as-measured data show the deformation to be 0.09-0.11 nm *rms*/°C. Finally, rigid-body segment positions show a relatively large average change in piston position with temperature, 10-24 nm/°C. For nearly all applications, the effect of the entire array moving upward with temperature is insignificant.

# 7. FUTURE WORK

Improving the segmented MEMS DM performance is an ongoing effort at Iris AO. From the analysis here, it is clear that additional improvements to the DM packaging would result in better temperature stability. Further, having the capability to calibrate the DM at the mean operating temperature or being able to apply temperature compensation to the DM positions would be beneficial for on-sky operation, particularly for cryogenic instruments.